# Temperature and Velocity Characteristics of Rotating Turbulent Boundary Layers Under Non-Isothermal Conditions

Zhi Tao (陶智)[1,2,3], Ruquan You (由儒全)[1,2], Yao Ma (马遥)[1,3], Haiwang Li (李海旺)[1,2,*]

[1)] *Research Institute of Aero-Engine, Beihang University Beijing, 100191, China*

[2)] *National Key Laboratory of Science and Technology on Aero Engines Aero-thermodynamics, Beihang University Beijing, 100191, China*

[3)] *School of Energy and Power Engineering, Beihang University Beijing, 100191, China*

(*Corresponding Author: 09620@buaa.edu.cn)

(Dated: 8 March 2022)

This paper describes an experimental investigation, by means of hot-wire anemometry, of the characteristics of velocity and temperature in a rotating turbulent boundary layer under isothermal and non-isothermal conditions. The ranges of experimental parameters are: Reynolds number from 10000 to 25000, rotational speed from 0 to 150 rpm, and $y^+$ from 1.8 to 100. The relative temperature difference is held constant at 0.1. Detailed velocity and temperature distributions in the boundary layer are measured in the rotating state, and a new criterion for boundary layer segmentation under rotation is proposed. The applicability of boundary layer theory under the rotating state is extended. The influence of Coriolis force and buoyancy on the velocity and temperature distributions in the turbulent boundary layers are analyzed. Coriolis force is found to play an important role in the behavior of the boundary layer under rotation, as it shifts the velocity and temperature boundary layers. Under isothermal conditions, such effects can be classified according to the dominant force: viscous, Coriolis, or inertial. Under non-isothermal conditions, buoyancy occurs. The buoyancy induced by the Coriolis force suppresses the effect of the Coriolis force, and the suppression effect increases with temperature difference. The variation of turbulent Prandtl number $Pr_t$ under rotation is also obtained.

## Nomenclature

$c_p$      specific heat

| Symbol | Description |
|---|---|
| $\vec{f}_{cen}$ | centrifugal force |
| $\vec{f}_{Cor}$ | Coriolis force |
| $\vec{f}_{buo,cen}$ | buoyancy induced by centrifugal force |
| $\vec{f}_{buo,Cor}$ | buoyancy induced by Coriolis force |
| $k_1$ | constant |
| $k_2$ | constant |
| $k_3$ | constant |
| Re | Reynolds number |
| Pr | Prandtl number |
| $Pr_t$ | turbulent Prandtl number |
| $q_w$ | wall heat flux |
| $\vec{r}$ | rotating radius |
| $T$ | temperature |
| $T_\infty$ | main stream temperature |
| $T_w$ | wall temperature |
| $T_\tau$ | wall friction temperature |
| $T^+$ | non-dimensional $T$ |
| $\vec{U}$ | velocity |
| $u$ | velocity in the x-direction |
| $u_\tau$ | wall friction velocity |
| $u^+$ | non-dimensional $u$ |
| $y$ | distance from the wall |
| $y^+$ | non-dimensional distance from the wall |
| $\beta$ | volume expansion coefficient |
| $\delta$ | thickness related to boundary layer |
| $\tau_w$ | wall shear stress |
| $\bar{\Omega}$ | rotational speed |
| $\rho$ | density |
| $\rho_0$ | density of main flow |
| $\lambda$ | thermal conductivity |
| $\mu$ | dynamic viscosity |
| $\nu$ | kinematic viscosity |
| $\varepsilon_m$ | turbulent momentum viscosity |
| $\varepsilon_h$ | turbulent thermal viscosity |

## I. BACKGROUND

### A. Introduction

The theory of the boundary layer proposed by Ludwig Prandtl greatly advanced the development of fluid mechanics and heat transfer, and Prandtl's logarithmic law has been widely used in the study of fluid mechanics. The classical theory of the boundary layer was developed, however, based on experiments in the non-rotating static state. Whether the classical theory of boundary layer is applicable to rotating fluid, and how the logarithm law changes under rotation, remain open research issues, and are crucial to the development of rotating machinery for aero engines and other engineering applications.

In 1962, Hill and Moon [1] conducted the first measurements of the velocity profile of the main stream in a smooth channel under a rotating condition, and found that the velocity distribution under a rotating condition is very different from that in a static state. However, limited by the testing facilities and measurement techniques of the time, that and subsequent measurements, including the one by Wagner and Velkoff, [2] were restricted to the main stream. No detailed velocity distribution in the boundary layer was obtained.

More recent researchers have adopted a variety of experimental techniques, such as PIV, [3] LDV, [4] and hot-wire anemometry, [5] to measure the flow field in a rotating channel. However, PIV and LDV have two disadvantages for use in rotating channels. First, since both methods use lasers, there will be laser reflection or scattering near the wall, meaning that the flow field near the wall cannot be accurately measured. Second, due to the large volume of equipment required for PIV and LDV, the apparatus cannot rotate with the experimental channel, so the relative velocity of the fluid can only be calculated by measuring the absolute velocity of the fluid in the rotating channel minus the migration velocity at the measured position. Di [3] pointed out that this method will result in much higher uncertainties than direct measurements, especially for flows with lower relative velocities in the boundary layer. Hot wire measurement does not require a laser and involves less bulky apparatus. Furthermore, hot wire measurement offers high spatial and temporal resolution, so it is suitable for the measurement of wall turbulence.

It was not until 1979 that Koyama, Masuda, et al. [5] began to study the velocity characteristics of rotating boundary layers. They used hot-wire anemometry and pitot probes to measure the velocity profiles of turbulent boundary layers along the spanwise centreline in a smooth straight channel at a fixed location under rotation. The thickness of the boundary layer and the wall friction factor were determined by analysing the mainstream velocity profile. The velocity profile in the boundary layer is significantly different under rotating conditions, and the wall friction factor is larger than in a non-rotating flow.

From 1985 to 2000, Joubert [6,7,8] and his team studied the turbulent boundary layer under the rotating state using hot-wire anemometers. Measurements of the velocity profiles (minimum $y^+ \approx 3$) in turbulent boundary layers were performed along the spanwise centreline in a smooth straight channel at different streamwise locations under the rotating state. The boundary layer thickness and wall friction factor were obtained. The viscous layer was found to be very little affected by rotation but the logarithmic region required a correction for rotation. Joubert's rotation correction method does not change the basic form of the logarithm law, changing only the empirical coefficient. Different from the conclusions of Koyama and Masuda et al. [5], the wall friction factor was found to decrease monotonically along the flow direction. At the same Reynolds number, the wall friction factor of the co-Coriolis surface is greater than its counterpart of the static wall, while that of the static wall is greater than that of the counter-Coriolis surface.

Nakabayashi and Kitoh [9] studied low-Reynolds-number turbulent boundary layers under rotation using hot-wire anemometers. The boundary layer in the rotating channel was divided into five sections: the viscous layer, the buffer layer, the logarithmic region, the Coriolis region, and the core region. Several length scales were identified to characterize the boundary layer: hydraulic diameter of the channel, thickness of the boundary layer, and thickness of the Coriolis force. The viscous and buffer layers are influenced only by viscosity, and the logarithmic region vanishes under severe rotation.

In sum, there has been little published work on the turbulent boundary layer in rotating flows, and no general conclusions have been reached regarding the flow characteristics of rotating boundary layers. Some previous observations appear to contradict each other, further constraining the understanding of the flow and thermal behaviours of rotating boundary layers. This lack of fundamental knowledge hinders, specifically, the development of the cooling technologies for rotary machinery. Most of the existing studies have been limited to isothermal flows, with attention focused on the flow behaviors of turbulent velocity boundary layers. Under non-isothermal conditions, however, temperature boundary layers appear, and play a crucial role in the process of heat transfer. A general question thus arises as to how the temperature boundary layer behaves under rotating conditions, and how it interacts with the velocity boundary layer. For the temperature boundary layer, the turbulent Prandtl number $Pr_t$ dictates the relative significance of momentum and heat transfer. For stationary flat turbulent air flows, $Pr_t$ generally falls in the range of 0.8-0.9. [10,11] The situation for rotating cases, however, is drastically different, with varying $Pr_t$.

The present experimental study has two goals. The first is to explore the underlying mechanisms of Coriolis force in determining the characteristics of turbulent velocity boundary layers under isothermal conditions. The second is to investigate the interactions between Coriolis and centrifugal forces, and the ensuing effect on turbulent velocity and temperature boundary layers in a non-isothermal flow field. The turbulent Prandtl number is quantified under the rotating state. The experimental conditions are Re = $10000 \sim 25000$, $\Omega$=0~150 rpm, $(T_w - T_\infty)/T_\infty = 0.1$, and $y^+ = 1.8 \sim 100$. Through this study, the mechanisms of the influence of the Coriolis force on the turbulent boundary layer are explained. Detailed velocity and temperature distributions in the boundary layer are measured in the rotating state, and a new criterion for boundary layer segmentation under rotation is proposed. The applicability of boundary layer theory under the rotating state is extended.

**B. Rotating test facility and test section**

Figure 1 shows schematically the experimental facility for measuring the characteristics of rotating flow. A detailed description of this facility has been given in our previous work, so it is described here only briefly [12] The facility has a vertical-axis design, and consists of a gas supply module, motor, slip ring, disk platform, data acquisition system, and test section modules. The maximum rotation radius and rotational speed are 1 m and 200 rpm, respectively. The measurement and diagnostic systems include hot wire, PIV, temperature liquid crystal, and auxiliary equipment for determining the flow and heat transfer behaviors.

A square cross-section channel was implemented to introduce a turbulent boundary layer flow under rotation, as shown in Fig. 2. Both clockwise and counter-clockwise rotations were considered, under different rotational forces. The length of the channel is 720 mm and the cross-sectional area is 80x80 mm square. A trip wire was installed at the entrance of the channel to facilitate laminar-to-turbulent transition near the surface and achieve fully developed turbulent flow.

To maintain constant heat flux for the thermal boundary condition, ITO (Indium Tin Oxide) glass was placed on an inner surface of the channel to generate heat from electricity. The channel wall was made of plexiglass with 15mm thickness, which provided thermal insulation. The Fourier law, Eq (1), is used to determine the heat flux on the wall.

$$q_w = -\lambda \frac{\partial T}{\partial y}\bigg|_{y \to 0} = -\lambda_n \frac{T_{n+1} - T_{n-1}}{2\delta} \qquad (1)$$

where the $\lambda$ is the local heat conductivity. With the neglect of radiation, the heat conduction normal to the wall is assumed to equal to the heat convection between the ITO and fluid, as shown in Fig 3.

The heat flux at the wall can be obtained through the measurement of the temperature gradient. It should be noted that the hot-wire probe cannot be placed very close to the wall due to the wall effect. [13] In the present work, the temperature distribution is almost linear in the near-wall region. The temperature gradient at $y^+ \approx 4$ is thus used to substitute the value at the wall. The method of determining the wall heat flux appears to be reliable.

## C. Rotating hotwire technique

In the present study, a 55P15 boundary layer probe (Dantec), as shown in Fig. 4, was used to measure the velocity in the rotating turbulent boundary layer. A modified 55P71 parallel probe was used to simultaneously measure the velocity and temperature boundary layers under the rotating state. In order to reduce the influence of rotation on signal transmission, the 54T42 mini CTA module (Dantec) and the temperature data acquisition module were placed on the rotating section. The measured hot-wire signal was transmitted in digital form to a stationary terminal through a USB slip ring. The temporal resolution of velocity measurement was up to 10 kHz, while the resolution for the simultaneous temperature and velocity measurements was limited by the response of the temperature probe, around 2 kHz [14] The hot wire and its supporting rod were fixed on a displacement mechanism. During the experiment, the hotwire was first placed at an initial position, where the distance between the hot wire and the wall was $y_0$. The measurement point was then moved $\Delta y$ by the displacement mechanism. The spatial resolution of the displacement mechanism was ±2 μm.

Hutchins and Choi [14] pointed out that in the linear sub-layer region, especially for $y^+$<3.5, there exists a wall effect of hot-wire due to aerodynamic blockage and direct conductive heat transfer. In this area, as the probe moves to the wall, the measured velocity increases, instead of decreasing to zero. The hot-wire probe overestimates the velocity magnitude. The same effect also occurs for temperature measurements. Figure 5 shows that in the region of $y^+$> 3.5, the wall effect of hot-wire vanishes. The measurements are thus valid and accurate, since $y^+$ is always larger than 3.5 in the present work.

The distance $y$ between the hot-wire probe and the solid wall is a key parameter in determining the measurement accuracy, due to vibration and centrifugal force when the apparatus is in use. A real-time positioning technique for the boundary layer hot-wire probe was developed to determine the initial position $y_0$ of the hot wire probe.

Figure 6 shows the method of dynamic positioning of $y_0$. First, the hot-wire probe was located at a specified initial position $y_0'$ near the wall. Second, the probe was moved away from the wall step by step to obtain a series of positions $(y_1, y_2,...)$ and velocities $(u_1, u_2, ...)$. Third, valid data in the linear sub-layer were chosen to fit the correlation of $u$ and $y$. Finally, because the line fit in sub-layer should pass through the origin, the intercept $\Delta y_0'$ was added to get the true initial position $y_0 = y_0' + \Delta y_0'$. The measurement accuracy of $y_0$ was ± 2 μm.

## II. Uncertainty Analysis

The velocity $u$ was measured by a hot-wire anemometer system. The uncertainty in the experimental data is mainly associated with calibration and measurement. In the present work, the Dantec Streamline Pro Automatic Calibrator was employed with an uncertainty of ± 1% or ± 0.02 m/s. A fourth-order polynomial curve was used to fit the calibration line, giving rise to an uncertainty of ±0.01 m/s. Measurement uncertainty is caused by the temperature difference between the fluid in the calibrator and that in the testing rotating channel. In the experiment, the fluid temperature in the channel increased slightly due to the heat released by the blower. Temperature correction was used to reduce this measurement uncertainty, as discussed in our previous work[12]. However, when the fluid temperature changes, the fluid density also changes, and this cannot be modified or neglected. This uncertainty can be expressed in terms of $\Delta T / (\sqrt{3} T)$, where $T$ is the fluid temperature in the calibrator. $\Delta T$ represents the fluid temperature difference between the calibrator and the rotating channel. In the present study, the maximum $\Delta T$ is 5 K, and $T$ is 300 K. The measurement uncertainty thus is about ± 1%. The overall uncertainty of the velocity measured with the hot wire anemometry system is ± 2% or ± 0.03 m/s. Uncertainty of measurement location ($y$) comes from the step movement $\Delta y$ of the hot wire probe driven by the motor. The positioning accuracy of the drive motor is ± 2 μm, thereby giving the uncertainty of the measurement location is ± 2 μm.

With the two basic variables ($u$ and $y$), all the other parameters can be calculated and their corresponding uncertainties can be estimated with the error transfer function. The friction velocity ($u_\tau$) is an important parameter, defined as:

$$u_\tau = \sqrt{\frac{\tau_w}{\rho}} = \sqrt{\left(\frac{\mu}{\rho}\frac{\partial u}{\partial y}\right)_w} \quad (2)$$

Its uncertainty is estimated to be in the range of 9.4-12.3%.

The main sources of uncertainty in the present analysis are associated with the test device, conditional control, data fitting, and regression. The directly measured quantities are temperature and velocity. Calibration showed that the uncertainty of the Omega T-type thermocouple which monitored the temperature of the nozzle outlet is about ±0.2 °C. Since the thermocouple cannot be spatially fully coincident with the hot-wire probe, the resulting control uncertainty is approximately ± 0.2 °C. The data fitting uncertainty can be divided into two parts: the surface fitting uncertainty of the calibration data and the fitting uncertainty caused by the spatial distance between the two hot-wire probes. The total uncertainty caused by the above two sources is ±0.2 °C. With consideration of all uncertainties discussed above, the overall uncertainty of the temperature measured with the hot wire anemometry system is ± 0.6 °C.

The measurement uncertainty of the airflow velocity at the nozzle outlet is ± 1% or ± 0.02 m/s according to the specification of the device. The surface fitting uncertainty of the velocity calibration data is ± 0.02 m/s, and the relative uncertainty of velocity measurement caused by temperature-induced density change is $\Delta T / (\sqrt{3} \times 273 \text{ K})$. Thus, the uncertainty is about ± 0.2 %. The total uncertainty of the velocity data is ± 1.2% or ± 0.04 m/s.

It is convenient to normalize flow quantities by wall units for data analysis and correlation. The wall quantities employed for parameter normalization are the friction velocity $u_\tau$ and friction temperature $T_\tau$, calculated as follows.

$$T_\tau = \frac{q_w}{\rho c_p u_\tau} \quad (3)$$

In Eq. (3), constant heat flow boundary conditions are adopted and the heat flux $q_w$ is about 380 W/m². This value may change under different experimental conditions. The measurement uncertainty of heat flux is about 6.4%. The uncertainties of these two parameters are calculated as follows. Since variable physical properties are used in the calculation, the uncertainty generated is relatively small and can be neglected.

$$k = \left(\frac{\partial u}{\partial y}\right)_w \quad (4)$$

$$\frac{\Delta u_\tau}{u_\tau} = \left(\sqrt{\left(\frac{\partial u_\tau}{\partial k} \Delta k\right)^2}\right) / u_\tau = \frac{1}{2} \cdot \frac{\Delta k}{k} = 4.5\% \quad (5)$$

$$\frac{\Delta T_\tau}{T_\tau} = \left(\sqrt{\left(\frac{\partial T_\tau}{\partial q_w} \Delta q_w\right)^2 + \left(\frac{\partial T_\tau}{\partial u_\tau} \Delta u_\tau\right)^2}\right) / T_\tau = 7.8\% \quad (6)$$

The dimensionless velocity and temperature can thus be determined as follows. The uncertainties of measured and deduced flow quantities are listed in Table 1.

$$y^+ = \frac{y u_\tau}{\nu}, \quad u^+ = \frac{u}{u_\tau}, \quad T^+ = \frac{T_w - T}{T_\tau} \quad (7)$$

$$\frac{\Delta y^+}{y^+} = \left(\sqrt{\left(\frac{\partial y^+}{\partial y}\Delta y\right)^2 + \left(\frac{\partial y^+}{\partial u_\tau}\Delta u_\tau\right)^2}\right) / y^+ = 4.6\% \qquad (8)$$

$$\frac{\Delta u^+}{u^+} = \left(\sqrt{\left(\frac{\partial u^+}{\partial u}\Delta u\right)^2 + \left(\frac{\partial u^+}{\partial u_\tau}\Delta u_\tau\right)^2}\right) / u^+ = 4.9\% \qquad (9)$$

$$\frac{\Delta T^+}{T^+} = \left(\sqrt{\left(\frac{\partial T^+}{\partial T}\Delta T\right)^2 + \left(\frac{\partial T^+}{\partial T_\tau}\Delta T_\tau\right)^2}\right) / T^+ = 8.8\% \qquad (10)$$

### III. Experiment results and discussion

#### A. Measurement validation

To validate the measurement accuracy, the measured turbulent velocity and temperature boundary layer data for stationary air flow over a flat plate were compared against the published in the literature. [7,11] Good agreement was achieved, as shown in Fig. 7.

The measured velocity profile of the isothermal velocity boundary layer closely matches published data[8]. The classical velocity logarithmic law is also matched, with a difference less than ± 2.5%. For the non-isothermal temperature boundary layer, the same level of agreement with the published data and the classical temperature logarithmic law is obtained. The maximum difference is less than ±4.5%. Both the velocity and temperature measurements appear to be highly credible.

#### B. Characteristics of turbulent velocity boundary layer under rotation

The turbulent velocity boundary layer under both isothermal and non-isothermal conditions were studied. In the latter situation, temperature gradients exist in the boundary layer and generate density gradients. In addition, rotation produces additional forces on the boundary layer, a phenomenon which will be described in the following section.

#### (a) Isothermal turbulent velocity boundary layer

Figure 8 shows the dimensionless velocity profiles of the isothermal turbulent boundary layers under different rotation conditions. The Coriolis force affects the velocity distribution in the boundary layer, as compared with the situation in the stationary state.

The magnitude of the Coriolis force in the boundary layer approximately equals the product of rotational speed and streamwise velocity, $f_{Cor} \approx 2\rho\Omega U$. According to the classical characteristics of the velocity distribution, Coriolis force along the wall normal direction is estimated as follows

$$\frac{f_{Cor}}{2\rho\Omega u_\tau} \approx u^+ = \begin{cases} y^+ & \text{sub-layer} \\ \frac{1}{\kappa}\ln y^+ + B & \text{log-layer} \end{cases} \qquad (11)$$

Coriolis acceleration is the derivative of velocity. Therefore, the magnitude of Coriolis effect is estimated as the integral of Coriolis force along the wall normal direction:

$$\int_0^{y^+} u^+ dy^+ \approx \begin{cases} \frac{1}{2}y^+ y^+ & \text{sub-layer} \\ \left(\frac{1}{\kappa}\ln y^+ + B - \frac{1}{\kappa}\right)y^+ & \text{log-layer} \end{cases} \qquad (12)$$

It is obvious that the order of Coriolis force in the linear sub-layer is at the level of $(y^+)^2$. Although it is parabolic, the distance from the wall is very small and the effect is negligible. In the logarithmic region (log-layer), the order of Coriolis force is at the level of $y^+$. Considering the distance from the wall, the effect of the Coriolis force is significant.

In the sub-layer, it can be estimated as

$$\nu\frac{\partial u}{\partial y} \approx \frac{\tau_w}{\rho} + \left(\frac{1}{\rho}\frac{\partial p}{\partial x} - \Omega^2 r_x\right)y - 2\nu\Omega y \qquad (13)$$

where the $r_x$ denotes the projection of $\vec{r}$ in the streamwise direction. Under the influence of rotation, $\tau_w$ changes less than 20%,[15] and the second term is approximately zero. Therefore, the Coriolis effect of the gradient of $u$ is at the level of $y^+$, and proportional to $\Omega$, which also means that with an increase of $\Omega$ the region affected by Coriolis will expand to the wall.

This phenomenon is commonly known as the Coriolis shift. The direction of shift depends on the projection of the Coriolis force along the wall normal direction, $\vec{f}_{Cor} = -2\rho_0\vec{\Omega}\times\vec{U}$ where $\rho_0$ represents the density of cold air, $\vec{\Omega}$ the rotational speed, and $\vec{U}$ the mainstream velocity. When $\Omega$ is greater than 0, the Coriolis force is opposite to the wall normal direction. The Coriolis force pushes the boundary layer toward the wall and suppresses the development of the boundary layer, rendering $u^+$ less than that in the non-rotating state at the same $y^+$ (shown by the red symbol in Fig. 9). When $\Omega$ is less than 0, the Coriolis force is aligned with the wall normal direction. It tends to expand the boundary layer away from the wall and promotes the development of the boundary layer, rendering $u^+$ larger than that in the non-rotating state at the same $y^+$ (shown by the blue symbol in Fig.9). The magnitude of the Coriolis shift depends on the Coriolis force. As shown in Fig. 9, the shift of velocity caused by the Coriolis force increases with the increase of $\Omega$. It is because $f_{Cor}$ increases with the increase of $\Omega$, which results the increase of the Coriolis shift.

In a rotating flow, the boundary layer can be divided into a sub-layer, a buffer layer, and the logarithmic region, and the Coriolis force is different in each region. When Ω is greater than 0, the boundary layer is suppressed by the Coriolis force and needs more space to develop, so that the sub-layer and logarithmic region are elongated, while the buffer layer is shortened. When Ω is less than 0, the Coriolis force pushes the boundary layer away from the wall to develop more slowly, so the sub-layer and logarithmic region are shortened as shown in Fig. 10.

According to the classical theory, a boundary layer forms when fluid flows over a solid wall, and its characteristics are determined by both viscous and inertial forces. In the main stream, the velocity gradient is relatively low, the inertial force is dominant, and the viscous force can be neglected. In a rotating state, besides viscous and inertial forces, Coriolis force plays an important role and must be taken into account. To characterize the dominant region of each force in a rotating boundary layer, a dimensionless analysis of the velocity field is carried out. The mainstream velocity is governed by the boundary layer thickness $\delta$, wall distance $y$, wall friction velocity $U_\tau$, kinetic viscosity $\nu$, and rotational speed Ω.

$$U = f_1(\delta, y, U_\tau, \nu, \Omega) \qquad (14)$$

The corresponding non-dimensional form is

$$U^+ = f_2(yU_\tau/\nu, U_\tau^2/\Omega\nu, U_\tau\delta/\nu) \qquad (15)$$

where $yU_\tau/\nu$ is $y^+$, which essentially is the Reynolds number based on the wall friction velocity $U_\tau$ and the wall distance $y$. $U_\tau^2/\Omega\nu$ can be expressed as $(yU_\tau/\nu)/(y\Omega/U_\tau)$, which is the ratio of the Reynolds number $y^+$ to the rotation number $y\Omega/U_\tau$; $U_\tau\delta/\nu$ is the Reynolds number based on the wall friction velocity $U_\tau$ and the boundary thickness $\delta$. The three dimensionless parameters are chosen to describe the boundary-layer behaviour under rotation in terms of the viscous force, Coriolis force, and inertia force. The characteristic length for the viscous force is $\delta_\nu = \nu/U_\tau$, the wall distance $y$ for $y^+ = 1$. The characteristic length for the Coriolis force is $\delta_c = U_\tau/|\Omega|$, which is the wall distance $y$ when the rotation number $|\Omega|y/U_\tau$ becomes unity. The characteristic length for the inertial force is the boundary thickness $\delta$. All of these parameters are non-dimensional.

The analysis of the isothermal turbulent velocity boundary layer is based on the dominant forces in the different regions. In the viscous force dominated region: $y < k_1\delta_\nu (k_1 = 30 \sim 50)$. In the Coriolis force dominated region: $k_2\delta < y < k_3\delta (k_2 = 0.008 \sim 0.015, k_3 = 0.2 \sim 0.3)$. And in the inertia-force dominating region: $y > k_3\delta (k_3 = 0.2 \sim 0.3)$.

**(b) Non-isothermal turbulent velocity boundary layer**

In a non-isothermal boundary layer over a heated wall, temperature variation causes a fluid density gradient. Buoyancy is generated by both the centrifugal and Coriolis forces, $\vec{f}_{buo} = \rho_0 \beta \vec{\Omega} \times (\vec{\Omega} \times \vec{r}) \Delta T + 2\rho_0 \beta \vec{\Omega} \times \vec{U} \Delta T$, where $\rho_0$ denotes the density of the mainstream air, $\beta$ the volume expansion coefficient, $\vec{\Omega}$ the rotational speed, $\vec{r}$ the rotating radius, and $\Delta T$ the temperature difference between the wall $T_w$ and mainstream $T_\infty$. Both isothermal and non-isothermal turbulent velocity boundary layers are considered in the present study, and the result demonstrates the effect of buoyancy on the characteristics of the boundary layer.

Buoyancy can be induced by both centrifugal and Coriolis forces. In the case of radial outflow and cold fluid over a heated wall $(\Delta T > 0)$, the direction of buoyancy caused by the centrifugal and Coriolis forces is opposite to its original direction, and the velocity in the boundary layer becomes slower than that without buoyancy. Figure 11 shows velocity profiles for different conditions.

The relationships among the centrifugal, Coriolis, and buoyancy forces are further analysed as follows.

Coriolis force

$$\vec{f}_{Cor} = -2\rho_0 \vec{\Omega} \times \vec{U} \qquad (16)$$

Centrifugal force

$$\vec{f}_{cen} = -\rho_0 \vec{\Omega} \times (\vec{\Omega} \times \vec{r}) \qquad (17)$$

Buoyancy induced by centrifugal force

$$\vec{f}_{buo,cen} = \rho_0 \beta \vec{\Omega} \times (\vec{\Omega} \times \vec{r}) \Delta T \qquad (18)$$

Buoyancy induced by Coriolis force

$$\vec{f}_{buo,Cor} = 2\rho_0 \beta \vec{\Omega} \times \vec{U} \Delta T \qquad (19)$$

For ideal gas, $\beta = 1/T_\infty$ (where $T_\infty$ is the mainstream temperature). Within the scope of the present study, the relative variation of temperature $\Delta T/T$ is about 0.1. The buoyancy induced from the Coriolis is about 10% of that induced by the centrifugal forces. The effect of buoyancy on the Coriolis shift is about 10% weaker than in an isothermal boundary layer. This weakening effect enhances with increasing temperature difference, as shown in Fig. 11.

## C. Non-isothermal temperature boundary layer

A non-isothermal flow field has a temperature boundary layer. Compared with the velocity boundary layer, the temperature boundary layer is much less studied, and no study has been reported in the open literature on the turbulent temperature boundary layer under rotation. The present work appears to be the first of its kind in this regard. Figure 12 shows the measured temperature distributions under different rotating conditions. The temperature boundary layer behaves in a manner similar to the velocity boundary layer under the effect of the Coriolis force, and the direction of the shift of the temperature boundary layer depends on the Coriolis force. When $\Omega$ is greater than 0, that is, the Coriolis force is opposite to the normal direction of the wall. The boundary layer is compressed toward the wall, rendering $T^+$ less than that of the non-rotating counterpart at the same $y^+$ (red symbol). When $\Omega$ is less than 0 (blue symbol), the Coriolis force is aligned with the normal direction of the wall and thus expands the boundary layer away from the wall. As a result, at the same position of $y^+$, $T^+$ is larger than in the non-rotating case (blue symbol). The trend is opposite to the situation with positive $\Omega$.

It is also found that the magnitude of the Coriolis shift in the temperature boundary layer depends on the Coriolis force. As shown in the Fig. 13, when $\Omega$ is greater than 0 (red symbol), the temperature shift caused by the Coriolis force increases with increasing rotational speed. The behaviour is reversed for positive $\Omega$ (blue symbol).

Turbulent Prandtl number is a significant dimensionless parameter characterizing the relative significance of momentum to heat transfer in a turbulent boundary layer. Rotation changes the flow stability; when $\Omega$ is greater than 0, the stability decreases, the turbulence fluctuation increases, and the transport capacity of momentum and heat increases. Temperature, however, as a passive transport scalar, is less affected by rotation than momentum, and $Pr_t$ increases. Conversely, $Pr_t$ decreases when $\Omega$ is less than 0. This parameter cannot be directly measured, but can be determined as follows:

The heat flux at the wall is $q_w$,

$$\frac{q_w}{\rho c_p} = -\frac{k}{\rho c_p}\frac{d\overline{T}}{dy} + \overline{v'T'} \qquad (20)$$

Based on the eddy viscosity hypothesis

$$-\overline{v'T'} = \varepsilon_h \frac{d\overline{T}}{dy} \qquad (21)$$

$$\left(\frac{1}{Pr} + \frac{\varepsilon_h}{\nu}\right)\frac{dT^+}{dy^+} = 1 \qquad (22)$$

For the velocity boundary layer

$$\left(1 + \frac{\varepsilon_m}{\nu}\right)\frac{du^+}{dy^+} = 1 \qquad (23)$$

By assuming equal turbulent shear stress in the logarithmic region of the velocity boundary layer and ignoring the molecular

viscosity, (23) becomes

$$\frac{\varepsilon_m}{\nu}\frac{du^+}{dy^+} = 1 \qquad (24)$$

By assuming equal turbulent heat flux in the logarithmic region of the temperature boundary layer and ignoring the molecular heat conduction, (22) becomes

$$\frac{\varepsilon_h}{\nu}\frac{dT^+}{dy^+} = 1 \qquad (25)$$

Define $K_V$ and $K_T$ as the gradients of dimensionless velocity ($u^+$) and temperature ($T^+$), respectively. We have

$$\Pr_t = \frac{\varepsilon_m}{\varepsilon_h} = \frac{K_T}{K_V} \qquad (26)$$

The turbulent Prandtl number can thus be calculated from the profiles of mean temperature and velocity. Figure 14 shows the variation of turbulent Prandtl number with rotation. It increases with increasing rotational speed for positive Ω; when Ω=90 rpm, for example, it increases by about 10%. Coriolis force reduces the thickness of the boundary layer. The resultant increase of the velocity gradient strengthens turbulence fluctuation. Coriolis force also promotes the mixing of hot and cold fluids in the boundary layer, thus enhancing thermal diffusion in this region. The influence of Coriolis force is to affect the temperature boundary layer through its effect on the velocity boundary layer, and the influence is hysteretic, leading to an increase of the turbulent Prandtl number when Ω is greater than 0.

When Ω is less than zero, the situation is reversed, due to the expansion of the boundary layer from the wall by the Coriolis force. In addition, the reduced temperature gradient hinders heat diffusion. However, as a consequence of the hysteretic effect of Coriolis force, the turbulent Prandtl number decreases with increasing rotational speed when Ω is negative.

## IV. Conclusion

In the present study, hot-wire anemometry was used to explore the characteristics of the velocity and temperature fields in rotating isothermal and non-isothermal turbulent boundary layers. The experimental conditions covered the parameter ranges of Re = 10000~25000 and Ω = 0~150 rpm. The relative temperature difference $(T_w - T_\infty)/T_\infty$ is 0.1. The velocity and temperature distributions in the range of $y^+$ from 1.8 to 100 were measured. The main conclusions from this work follow.

(1) Different from the stationary isothermal boundary layer, the velocity distribution shifts due to the Coriolis force under rotation, and this shift increases with increasing Coriolis force.

(2) Three main forces act on the isothermal turbulent velocity boundary layer: the viscous force, Coriolis force, and inertial force. Under the action of these forces, the isothermal velocity boundary layer has distinct regions. A new criterion for division of the isothermal turbulent velocity boundary layer is proposed. I. Viscous force dominated region: $y < k_1 \delta_v, (k_1 = 30 \sim 50)$. II. Coriolis force dominated region: $k_2 \delta < y < k_3 \delta$, $(k_2 = 0.008 \sim 0.015, k_3 = 0.2 \sim 0.3)$. III. Inertial force dominated region: $y > k_3 \delta (k_3 = 0.2 \sim 0.3)$.

(3) In the non-isothermal turbulent boundary layer under rotation, buoyancy is generated (centrifugal force induced buoyancy force and Coriolis force induced buoyancy). Within the scope of this paper, the direction of the buoyancy is opposite to the original force, so the buoyancy induced by the Coriolis force weakens the velocity shift caused by Coriolis force. And the suppression strengthens with the increase of temperature difference.

(4) In the non-isothermal temperature boundary layer under the rotating state, Coriolis force causes a temperature boundary layer shift which strengthens with Coriolis force.

(5) Under rotation, Coriolis force affects $Pr_t$ through the velocity and temperature boundary layer. When Ω is greater than 0, $Pr_t$ increases with the increase of rotational speed. When Ω is less than 0, $Pr_t$ decreases with the increase of rotational speed.

## ACKNOWLEGMENTS

<>This work was supported by the National Natural Science Foundation of China [Grant Number 51906008]; Fundamental Research Funds for the Central Universities [Grant Number YWF-21-BJ-J-822]; and National Science and Technology Major Project [Grant Number 2017-III-0003-0027]. The funding sources were not involved in study design, in the collection, analysis and interpretation of data, in the writing of the report, or in the decision to submit the article for publication.

## DATA AVAILABILITY STATEMENT

The data in this manuscript are available from the corresponding author upon reasonable request.

**References**

[1] P. G. Hill, and I. M. Moon, Gas Turbine Laboratory Report, MIT (1962).

[2] R. E. Wagner, and H. R. Velkoff, Journal of Engineering for Power 94, 261 (1975).

[3] D. S. Alberto, T. Raf, and A. V. D. B. René, Experiments in Fluids 44, 179 (2008).

[4] T. M. Liou, and C. C. Chen, Journal of Turbomachinery 121, 167 (1999).

[5] H. Koyama, S. Masuda, I. Ariga, and I. Watanabe, Journal of Engineering for Power 101, 23 (1979).

[6] J. H. Watmuff, H. T. Witt, and P. N. Joubert, Journal of Fluid Mechanics 157, 405 (1985).

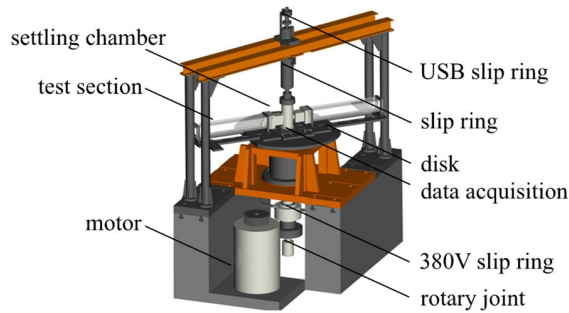

FIG. 1 Rotating test facility

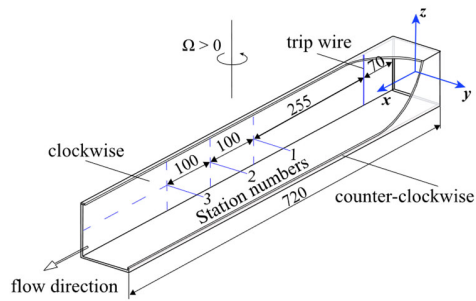

FIG. 2 Test channel

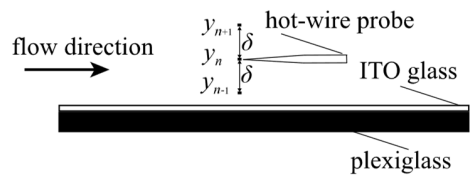

Fig. 3 The method of heat flux measurement

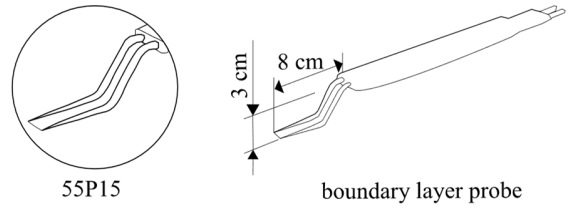

(a) The boundary layer probe

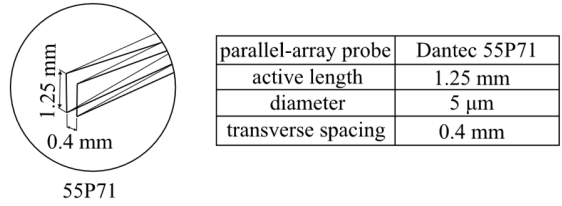

(b) The main parameters of the modified parallel-array probe

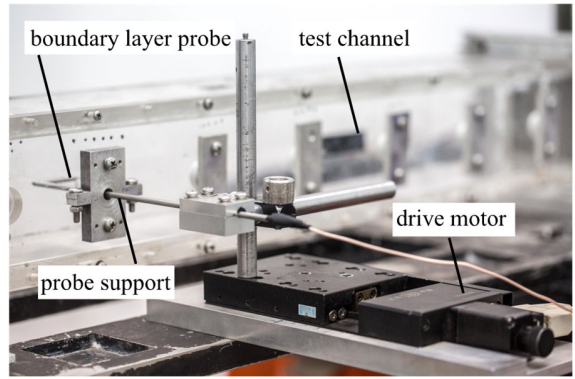

(c) The arrangement of the probe and its support mechanisms

FIG. 4 Hot wire probe and implementation

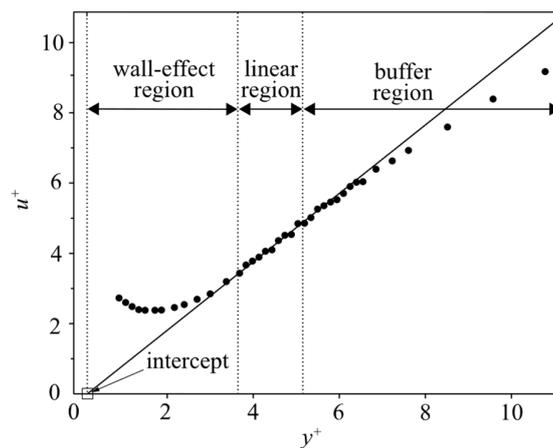

Fig. 5 Experimental data from a detailed inner region boundary layer traverse, including limits of linearity and wall-effect region. Both dimensional and non-dimensional axes are represented in the reference of 14

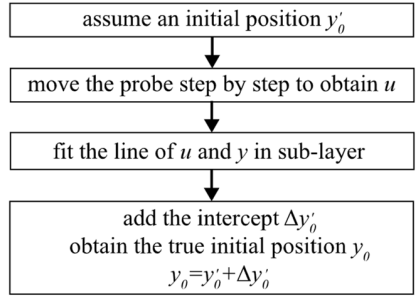

FIG. 6 Accurate positioning technique for hot wire probe in constrained space

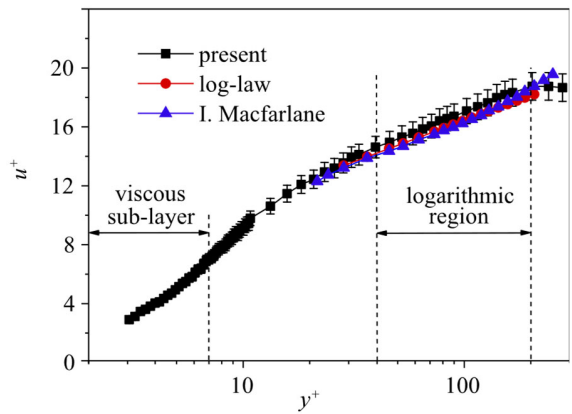

(a) Distribution of non-dimensional velocity

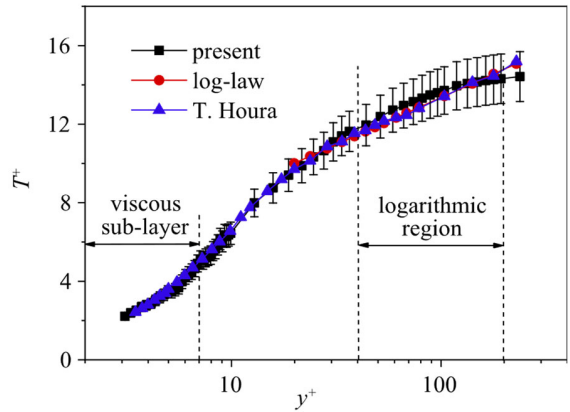

(b) Distribution of non-dimensional temperature

FIG. 7 Comparison with previous work (Macfarlane[7] and Houra[11])

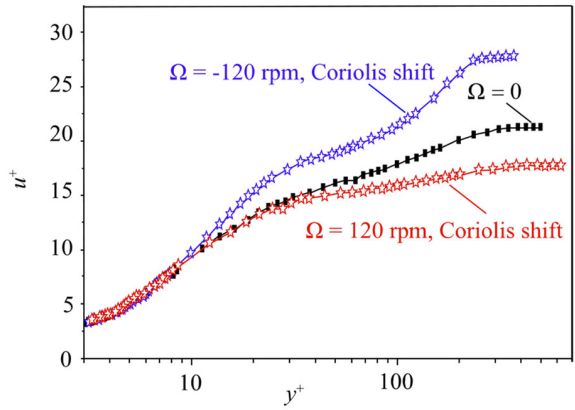

FIG. 8 Distribution of isothermal velocity boundary layer under Coriolis force

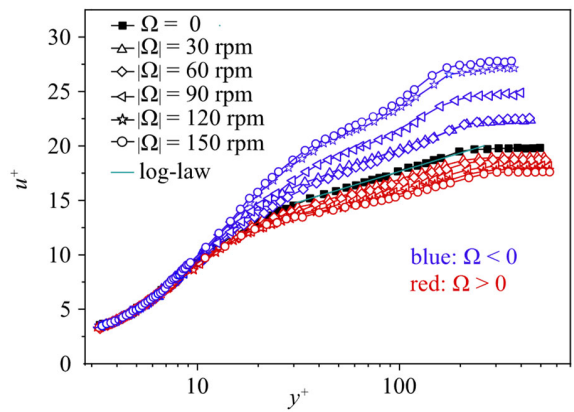

FIG. 9 Isothermal velocity boundary layer under different Coriolis forces

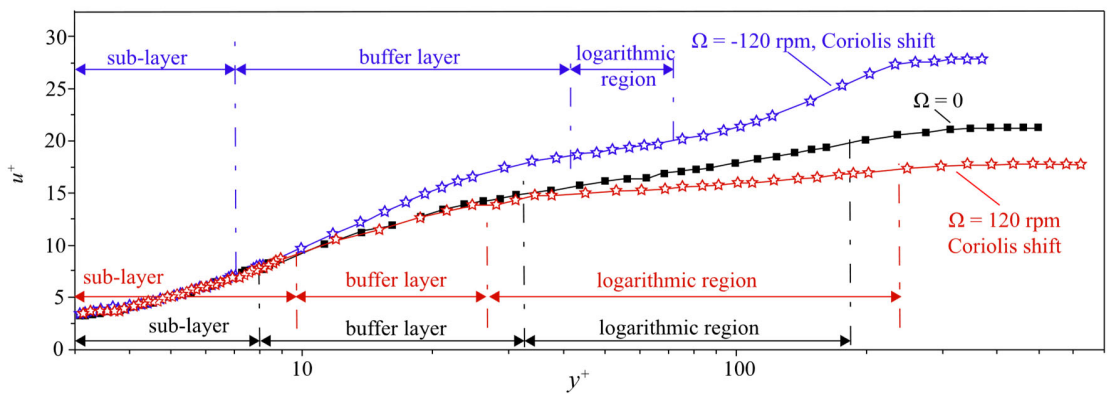

FIG. 10 Regions of turbulent velocity boundary layer under Coriolis force

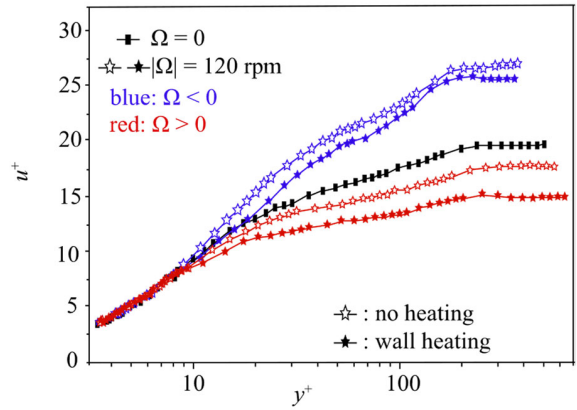

FIG. 11 Effect of buoyancy on non-isothermal turbulent velocity boundary layer

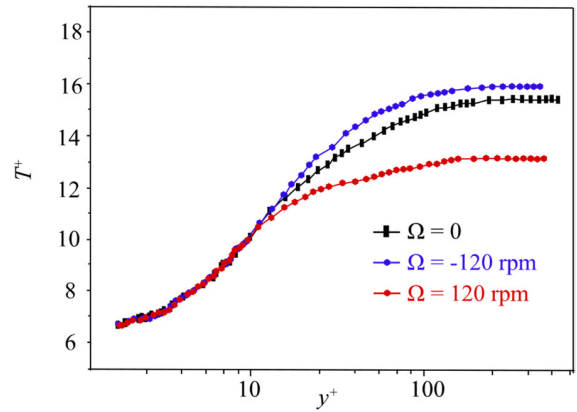

FIG. 12 Temperature distributions in non-isothermal turbulent temperature boundary layer

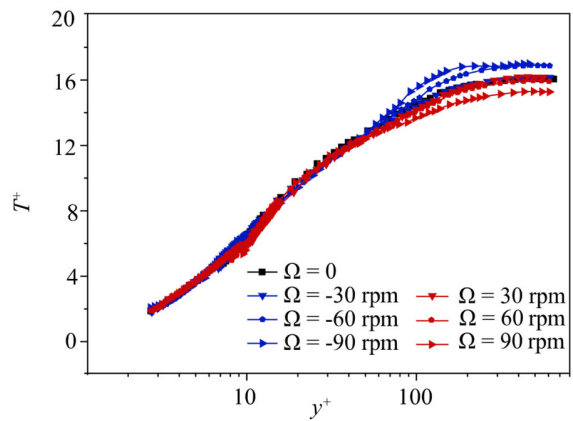

FIG. 13 Temperature distributions in non-isothermal boundary layers under rotation

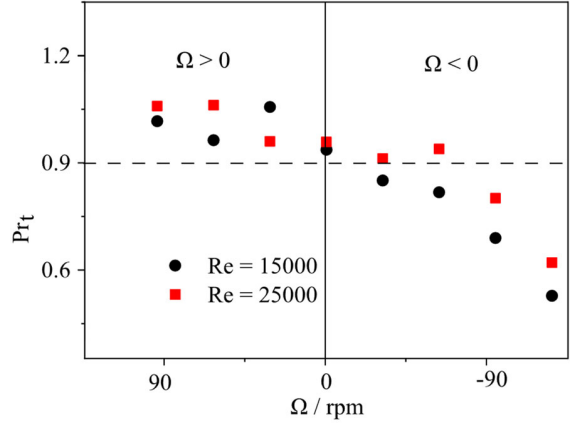

FIG. 14 Turbulent Prandtl number at different rotational speeds

TABLE. 1. Uncertainty of flow parameters

| Parameters | Uncertainty |
|---|---|
| $u_\tau$ | 4.5% |
| $T_\tau$ | 7.8% |
| $y^+$ | 4.6% |
| $u^+$ | 4.9% |
| $T^+$ | 8.8% |